\documentclass[cameraready]{Interspeech}

\title{Beyond Residual Connections: Manifold-Constrained Hyper-Connections for Robust Speaker Representation Learning}

\author[affiliation={1}, orcid=0009-0001-0828-2933, equalcontribution]{Zezhong}{Jin}
\author[affiliation={2}, orcid=0009-0008-7612-2051, equalcontribution]{Xiaoyu}{Wang}
\author[affiliation={3}, orcid=0000-0002-0519-7434,]{Zhe}{Li}
\author[affiliation={1}, orcid=0000-0003-2121-5935]{Chong-Xin}{Gan}
\author[affiliation={1}, orcid=0000-0003-0760-8350]{Zilong}{Huang}
\author[affiliation={1}, orcid=0000-0001-8854-3760, correspondingauthor]{Man-Wai}{Mak}
\author[affiliation={1}, orcid=0000-0001-9133-3000, correspondingauthor]{Kong Aik}{Lee}

\address{
    $^1$ Dept. of EEE, The Hong Kong Polytechnic University \quad $^2$ Baidu Inc. \quad $^3$ Speech, Language, and Cognition Laboratory, The University of Hong Kong, Hong Kong SAR
}

\email{zezhong.jin@connect.polyu.hk}


\usepackage{comment}
\usepackage{subcaption}
\usepackage{graphicx}
\usepackage{url}
\usepackage[table]{xcolor} 
\definecolor{myyellow}{rgb}{1, 1, 0.8} %
\definecolor{mygreen}{HTML}{E6F4EA}

\begin{document}

\maketitle

\begin{abstract}
Residual connections are fundamental to deep speaker recognition models, such as ECAPA-TDNN and ResNet. However, standard identity mapping limits information flow to a single path, constraining representation capacity. We introduce \textbf{M}anifold-Constrained \textbf{H}yper-\textbf{C}onnections (\textbf{mHC}), reformulating residual paths as a multi-stream evolution where information is mixed through a doubly stochastic matrix. By employing Sinkhorn-Knopp iterations, mHC ensures energy conservation by preserving signal intensity and feature mean, which stabilizes gradients and mitigates signal degradation in complex networks. We evaluate mHC by replacing standard residual connections in backbones including ECAPA-TDNN, ResNet-34, Res2Net, and E-Res2Net. Extensive experiments on VoxCeleb1 demonstrate that mHC connections consistently enhances performance across all architectures, highlighting its effectiveness for robust speaker representation learning.
\end{abstract}

\section{Introduction}
Speaker recognition has become an indispensable technology in various security-sensitive domains, including biometric authentication, secure financial transactions, and forensic analysis. In recent years, the rapid advancement of speaker recognition has been primarily driven by the development of powerful deep learning architectures \cite{he2016deep, desplanques2020ecapa, jin2026uncertainty, jin2025adversarially, jin24b_interspeech, jin2025denoising,li2025disentangling,jin2026distilling,fu2022vision, qin2025variational,jin2024self}. Among these, backbones such as ResNet \cite{he2016deep}, ECAPA-TDNN \cite{desplanques2020ecapa},  and the more recent E-Res2Net \cite{zhou2023eres2net} have demonstrated strong performance and robustness in extracting speaker-discriminative embeddings. The fundamental cornerstone of these architectures is the residual connections \cite{he2016deep}, which enable the training of deeper networks by establishing an identity mapping that mitigates the vanishing gradient problem.
\begin{figure*}[t]
  \centering
  
  \begin{subfigure}[b]{0.31\textwidth} 
    \centering
    \includegraphics[height=4.0cm, width=!]{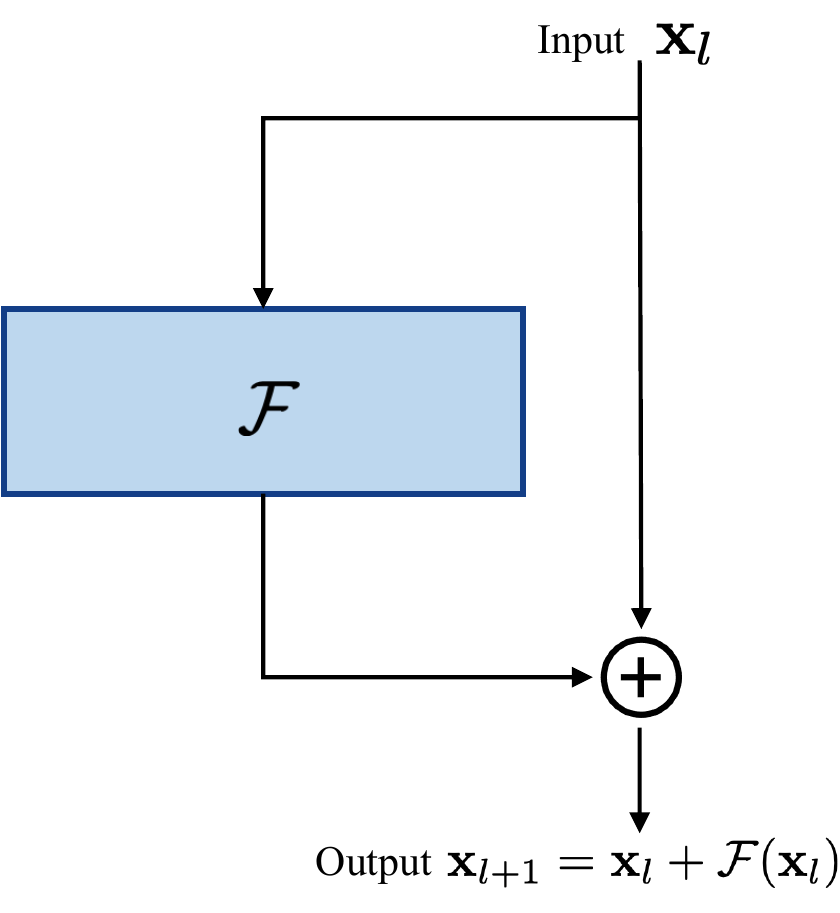} 
    \caption{Standard Residual Connection}
    \label{fig:residual}
  \end{subfigure}
  \hspace{0.001\textwidth}
  \begin{subfigure}[b]{0.31\textwidth}
    \centering
    \includegraphics[height=4.0cm, width=!]{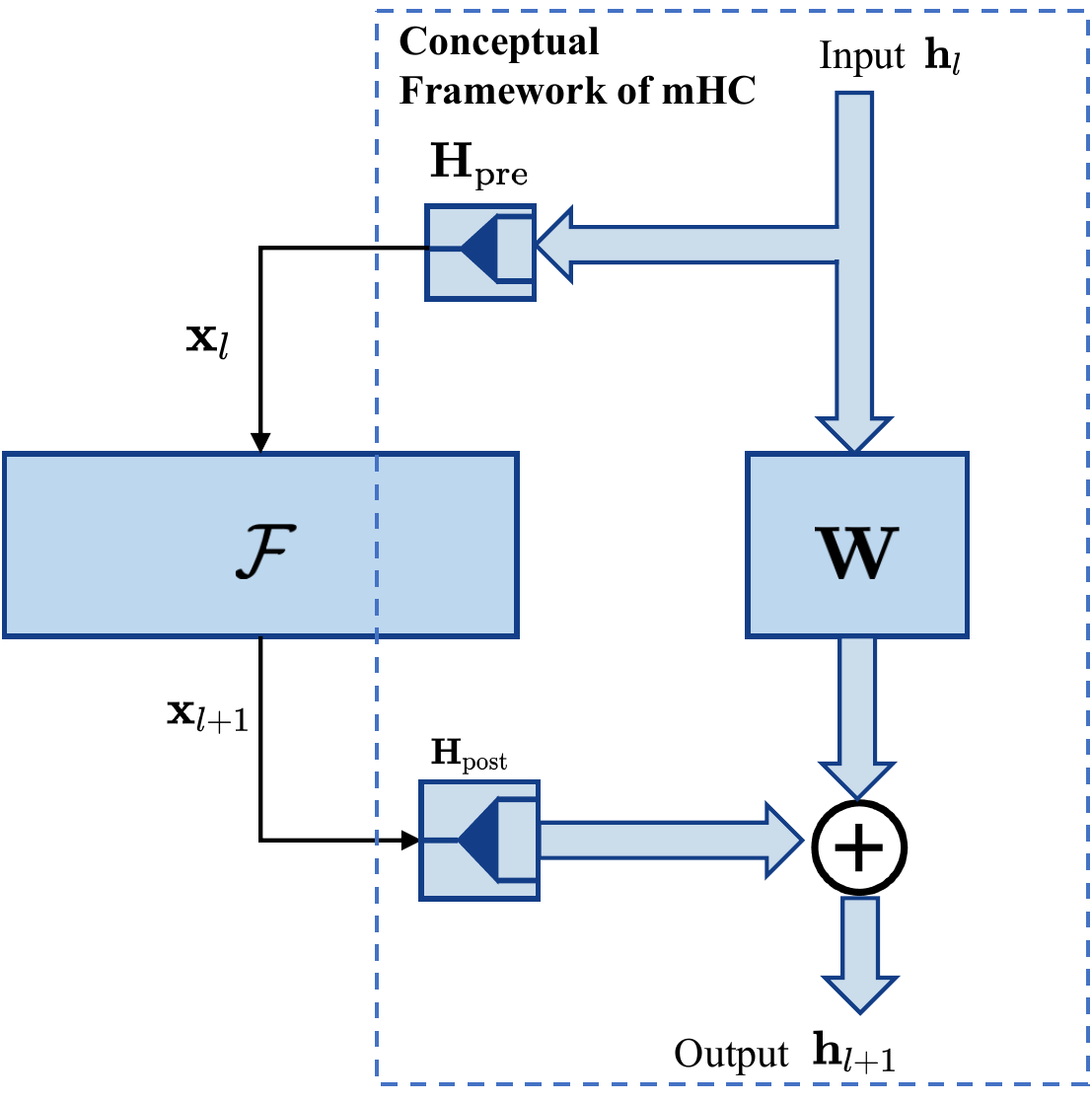} 
    \caption{High-level mHC}
    \label{fig:mhc}
  \end{subfigure}%
  \hspace{0.001\textwidth}
  \begin{subfigure}[b]{0.31\textwidth}
    \centering
    \includegraphics[height=4.0cm, width=!]{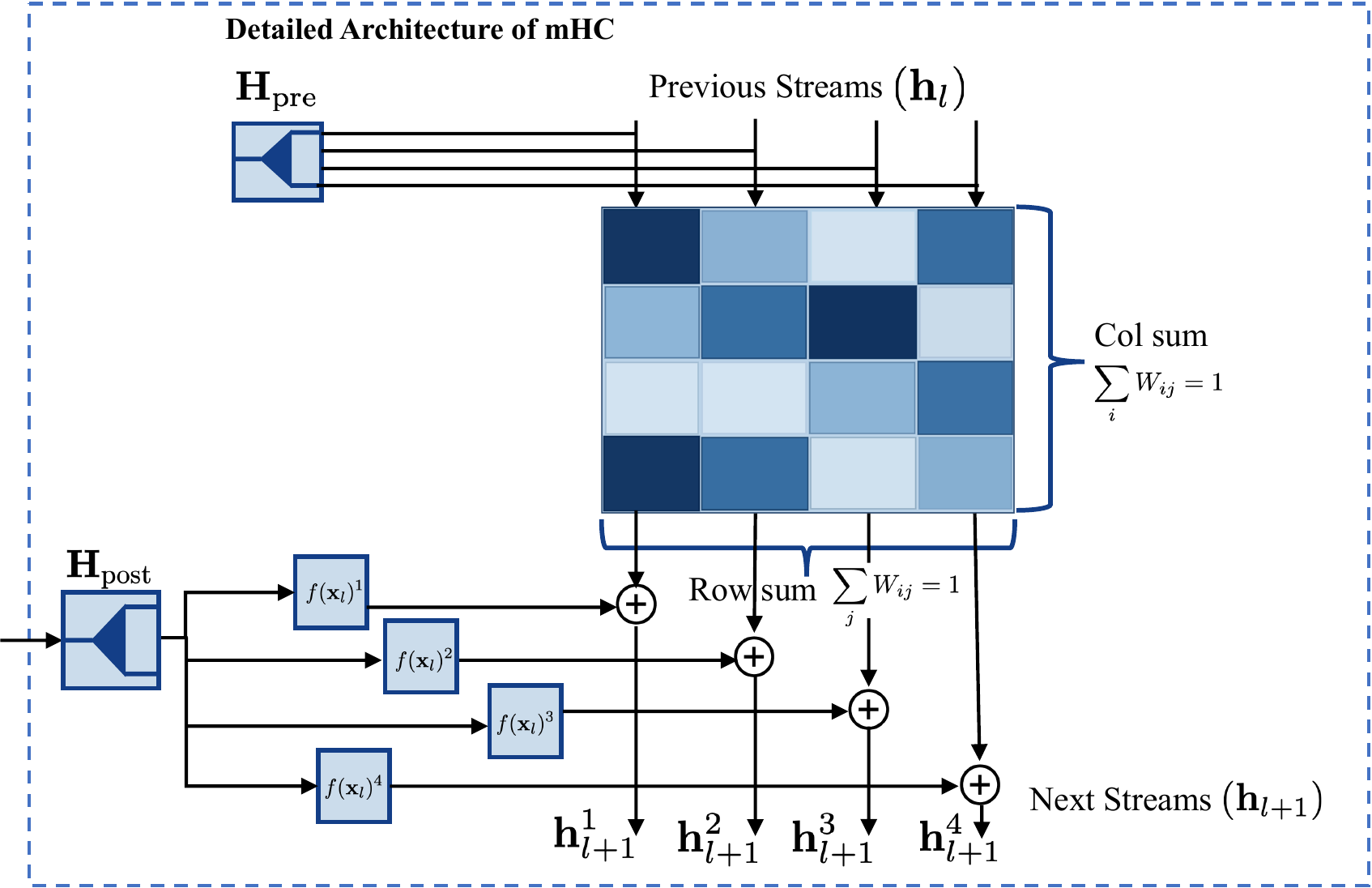}
    \caption{Detailed mHC}
    \label{fig:mhc_detail}
  \end{subfigure}

  \caption{Architectural comparison between the standard residual shortcut and mHC: (a) standard residual connection, (b) high-level mHC overview, and (c) detailed mHC design, where $\mathbf{H}_{\mathrm{pre}}$ aggregates streams before $\mathcal{F}(\cdot)$ and $\mathbf{H}_{\mathrm{post}}$ splits the output back into streams.}
  \label{fig:comparison}
\end{figure*}

Structurally, the identity mapping in standard residual learning relies on simple element-wise addition. In the formulation $\bold{y} = \bold{x} + f(\bold{x})$, each feature channel is essentially processed as an isolated path, where the $i$-th channel of the input $\bold{x}^{i}$ is strictly merged with the corresponding $i$-th channel of the transformation $f(\bold{x})^i$. This point-to-point summation lacks a mechanism for inter-channel information exchange, which may constrain the model's capacity \cite{gao2019res2net}. As the network grows deeper, this lack of interaction causes successive layers to produce highly correlated features, a phenomenon known as feature redundancy \cite{huang2017densely,cogswell2016reducingoverfittingdeepnetworks}. Without a dynamic protocol to ``mix" and ``refresh" information across channels, the model essentially repeats existing patterns rather than evolving more complex representations \cite{veit2016residual}. This limitation becomes increasingly evident in modern Transformer architectures \cite{vaswani2017attention,jin2025trink,brown2020gpt3,devlin2019bert,touvron2023llama, huang2024mm}, where model depth scales dramatically and the standard residual pathway can become less effective in maintaining stable and informative signal propagation \cite{bai2023residual, xie2025mhc,heddes2025deepcrossattention}. For tasks requiring high precision, this redundancy limits the model's discriminative resolution, making it difficult to distinguish between subtlely different inputs.

To address these limitations, Hyper-Connections (HC) was proposed in \cite{zhu2024hyper}, which expand the traditional single-path hidden state into multiple parallel streams by partitioning feature channels into several disjoint groups. In each layer, these streams are interactively redistributed through a learnable mixing matrix, allowing the model to aggregate and exchange information across different feature subspaces simultaneously. While this design encourages information flow, its critical limitation lies in the loss of the identity mapping property. In a standard ResNet, the shortcut connection acts as a fixed Identity Matrix ($I$), ensuring that the signal passes through unchanged. In contrast, HC replaces this fixed identity with unconstrained learnable matrices. Because these matrices are not forced to be identity-preserving, they cannot guarantee a consistent signal scale. Consequently, as the network depth increases, the signal magnitude tends to either explode or vanish, leading to severe training instability and preventing the model from converging. This wastes substantial training time and computational resources. To restore this essential stability, Xie et al.~\cite{xie2025mhc} introduce Manifold-Constrained Hyper-Connections (mHC), which imposes a manifold constraint on the mixing matrix to re-establish the identity-preserving property while retaining the high-bandwidth benefits of multi-stream interaction.

Unlike unconstrained Hyper-Connections (HC) that suffer from numerical instability, mHC reformulates information flow as a stable multi-stream process by employing Sinkhorn iterations \cite{sinkhorn1967concerning} to project the mixing matrix onto a doubly stochastic manifold, where both row and column sums are constrained to unity. This property ensures energy conservation across streams, providing a theoretically grounded safeguard against gradient explosion and vanishing \cite{he2016deep} while maintaining high-bandwidth communication. Regarding the implementation details, we propose a crucial optimization over the original framework: while the vanilla mHC relies on a computationally intensive dynamic mapping—where the mixing matrix is generated via high-dimensional projections of input features—we adopt an efficient static parameterization strategy. By employing a standalone learnable mixing matrix $\mathbf{W} \in \mathbb{R}^{n \times n}$ instead of an input-dependent transformation, the parameter overhead is effectively reduced from $\mathcal{O}(nCn^2)$ to $\mathcal{O}(n^2)$. This refinement is vital for lightweight backbones like ECAPA-TDNN, enabling the model to enjoy the benefits of multi-stream interaction with negligible computational cost.

In summary, this paper makes two primary contributions. First, to the best of our knowledge, this is the first work to introduce Manifold-Constrained Hyper-Connections (mHC) to the field of speaker recognition, redefining the conventional single-path residual connection as a stable, multi-stream information exchange protocol. Notably, this architectural upgrade achieves significant representational gains with a negligible increase in both parameter count and computational complexity (FLOPs), making it a highly efficient solution for deep embedding networks. Second, we conduct extensive experiments across multiple mainstream speaker embedding architectures and datasets to validate the robustness of the proposed method. The consistent performance improvements observed in various backbones, including ECAPA-TDNN, ResNet-34, Res2Net, and the E-Res2Net family, demonstrate that mHC serves as a universal and effective approach.

\section{Methodology}

\subsection{Manifold-Constrained Hyper-Connections}
Fig.~\ref{fig:residual} illustrates the standard residual connection. At layer $l$, the output is computed as
$\mathbf{x}_{l+1} = \mathbf{x}_l + \mathcal{F}(\mathbf{x}_l)$,
where $\mathbf{x}_l$ is passed through the shortcut unchanged and then added to the transformation block $\mathcal{F}(\mathbf{x}_l)$. This identity shortcut helps to train deep networks, but the shortcut itself is \emph{channel-wise independent}: the $c$-th channel of $\mathbf{x}_l$ is only added to the $c$-th channel of $\mathbf{x}_l$, so there is no explicit mechanism to redistribute information across channels within the skip path.

In Fig.~\ref{fig:mhc}, we use arrow widths to highlight the structural evolution: the thin arrows correspond to the single-path flow used in standard residual connections (Fig.~\ref{fig:residual}), while the wide arrows symbolize the multi-stream state of mHC, where information is processed in parallel subspaces. This visual distinction clarifies how mHC generalizes the rigid identity shortcut into a more expressive hyper-connection. In detailed mHC (Fig.~\ref{fig:mhc_detail}), the hidden state at layer $l$ is represented as $N$ parallel streams, $\mathbf{h}_{l} = \{\mathbf{h}_{l}^1, \mathbf{h}_{l}^2, \dots, \mathbf{h}_{l}^N\}$. These streams are first aggregated via an operator $\mathbf{H}_{\text{pre}}$ (concatenation) to form a unified feature map $\mathbf{x}_{l}$, which is then fed into the transformation block $\mathcal{F}(\cdot)$. The resulting output is subsequently partitioned by the operator $\mathbf{H}_{\text{post}}$ into $N$ disjoint streams $\{f(\mathbf{x}_l)^1, \dots, f(\mathbf{x}_l)^N\}$. This structure enables a stream-wise update where the transition from $\mathbf{h}_{l}$ to $\mathbf{h}_{l+1}$ is formalized as:

\begin{equation}
\mathbf{h}_{l+1}^i = \sum_{j=1}^{N} W_{ij}\,\mathbf{h}_{l}^j + f(\mathbf{x}_l)^i, \quad \forall i \in \{1, \dots, N\}.
\end{equation}

Inter-stream communication is governed by the learnable mixing matrix $\mathbf{W} \in \mathbb{R}^{N \times N}$, where $W_{ij}$ controls the information flow from the $j$-th historical stream to the $i$-th updated stream. This weighted summation provides the explicit mechanism for inter-channel communication missing in standard residuals. Unlike identity shortcuts where information is confined to isolated channels, $\mathbf{W}$ allows every stream $i$ to selectively aggregate and ``refresh" its content with all $N$ previous streams, transforming the shortcut into a high-bandwidth system for representation exchange.

To maintain numerical stability, $\mathbf{W}$ is projected onto a doubly stochastic manifold using Sinkhorn-Knopp iterations. We first construct a nonnegative matrix $\mathbf{A} = \exp(\mathbf{\Theta})$ from a learnable parameter matrix $\mathbf{\Theta}$. Then, we alternately apply row normalization $\mathbf{A} \leftarrow \text{diag}(\mathbf{A}\mathbf{1}_N)^{-1}\mathbf{A}$ followed by column normalization $\mathbf{A} \leftarrow \mathbf{A}\,\text{diag}(\mathbf{1}_N^\top\mathbf{A})^{-1}$. After a small number of iterations (typically $k=3$), we obtain the projection matrix $\mathbf{W}$ as the normalized matrix $\mathbf{A}$ that enforces unit row and column sums. This ``Energy Conservation" constraint preserves the signal scale across streams, preventing the signal from exploding or vanishing with depth. Compared with the standard residual shortcut, mHC enables active, learnable mixing while retaining the robust optimization properties of the original architecture.

\subsection{Integrated Block Architecture}
The mHC is designed as a modular and flexible component, allowing it to serve as a drop-in replacement for standard residual shortcuts across various neural architectures. By maintaining the original input and output dimensions of each block, mHC can be seamlessly integrated into existing backbones without altering their macro-structures or increasing the overall depth of the network.

For ResNet-34 and Res2Net, the networks are organized into four stages, each containing a sequence of residual blocks: the standard BasicBlocks (comprising two $3 \times 3$ convolutional layers) for ResNet-34, and the multi-scale Res2Blocks for Res2Net. Within each stage, we replace the identity shortcuts between successive blocks with mHC modules. To handle transitions between stages, we employ a dimension matching step where streams are aggregated and processed by a $1 \times 1$ convolution shortcut to match new resolutions and channel counts before being re-partitioned. This modification ensures that the multi-stream state is preserved and mixed as information flows through the stacked blocks of each stage.

In ECAPA-TDNN, the backbone consists of an initial 1D convolutional layer followed by a series of SE-Res2Blocks. Each SE-Res2Block contains a bottleneck structure: a $1\times1$ convolution for expansion, a multi-scale Res2Net layer for temporal processing, a $1\times1$ convolution for projection, and a Squeeze-and-Excitation (SE) attention module. We replace the main residual connection that skips over these internal operations with an mHC. By substituting this specific shortcut, we transform the local skip-connection of the SE-Res2Block into an active multi-stream interaction layer.

\section{Experiment Setting}

\subsection{Datasets}
\textbf{Training data.} We used the VoxCeleb2 development set \cite{chung2018voxceleb2} (5,994 speakers) for training. VoxCeleb2 is multilingual, while the majority of utterances are in English.

\textbf{Evaluation data.} We report results on the standard VoxCeleb1 test sets \cite{nagrani2017voxceleb}, namely Vox-O, Vox-E, and Vox-H. Vox-O contains 37,611 trials from 40 speakers; Vox-E contains 579,818 trials from 1,251 speakers; and Vox-H contains 550,894 trials from 1,190 speakers. We also evaluated system robustness on the VoxSRC 2021 validation set (VoxSRC21-val). Since the labels of the VoxSRC 2021 test set are not publicly available, we used VoxSRC21-val, which contains 60,000 evaluation trials sampled from VoxCeleb1.

\textbf{Data augmentation.} We followed the data augmentation strategy in Kaldi's recipes \cite{povey2011kaldi}. Specifically, we added noise, music, and babble to the training data using MUSAN \cite{musan2015} and created reverberated speech using the RIR dataset \cite{ko2017reverb}.
\subsection{Model Configurations}
We conducted experiments using four widely adopted speaker-embedding backbones of varying model sizes; their parameter counts are listed in Table~\ref{tab:voxceleb_results}.

\textit{ResNet34}: ResNet-34 \cite{he2016deep} is a residual CNN backbone that eases optimization via identity skip connections. It is typically organized into four stages of residual blocks, where each stage operates at a different channel width/resolution and the residual connections help preserve information flow in the deep network. Our ResNet-34 speaker encoder contains 6.34M parameters.

\textit{Res2Net}: Res2Net \cite{gao2019res2net} can be viewed as an enhanced residual design built on top of the ResNet idea; their key differenceis that each residual block in Res2Net explicitly models multi-scale features. Concretely, it splits feature channels into multiple groups and processes them in a hierarchical residual-like manner within a single block, enabling richer multi-scale representations without increasing depth. Our Res2Net speaker encoder contains 4.03M parameters.

\textit{ECAPA-TDNN-S}: ECAPA-TDNN \cite{desplanques2020ecapa} is a TDNN-based \cite{peddinti2015time} speaker encoder with enhanced channel attention and feature aggregation. The small model (ECAPA-S) uses a channel size of 512 and contains 6.19M parameters.

\textit{ECAPA-TDNN-L}: The large model (ECAPA-L) increases the channel size to 1024 and contains 20.76M parameters. Following the ECAPA-TDNN design \cite{desplanques2020ecapa}, it leverages squeeze-and-excitation style channel attention \cite{hu2018squeeze} and multi-layer feature aggregation to capture complementary temporal information.

The integration of mHC varies slightly across architectures. For ResNet-34 and Res2Net, we replaced the inter-block residual connections with mHCs. For ECAPA-TDNN, we replaced the residual connection inside the SE-Res2Block with an mHC.

\subsection{Training and Evaluation Protocol}
We trained all student speaker encoders using SGD with a batch size of 256. Each utterance was randomly cropped into 3 seconds and augmented with a probability of 0.8 during training. After augmentation, we extracted 80-dimensional filter bank (Fbank) features using a 25-ms frame length and a 10-ms frame shift, and we used the resulting Fbank features as input to the embedding network. We used a linear learning-rate warmup for the first 5 epochs, increasing the learning rate from $5\times10^{-5}$ to $0.2$, followed by a cosine decay schedule down to $5\times10^{-6}$. We optimized the models using AAM-Softmax loss \cite{deng2019arcface} with a scale of 32 and a margin scheduler: the margin was kept at 0.0 until epoch 20, then increased to 0.3, and fixed from epoch 50 onward.

Performance is reported using equal error rate (EER) and minimum detection cost function (minDCF), with $P_{\mathrm{target}}=0.01$. All experiments and evaluations were conducted using the 3D-Speaker toolkit \cite{chen20253d}.\footnote{\url{https://github.com/modelscope/3D-Speaker}}

\section{Experiment Results}

\subsection{Performance on VoxCeleb and VoxSRC21-val}

Table~\ref{tab:voxceleb_results} summarizes the results on the VoxCeleb1 test sets (VoxCeleb1-O/E/H). Replacing the standard residual shortcut with mHC consistently improves performance across all four backbones without increasing model parameters. For Res2Net, mHC reduces EER from 1.56\% to 1.41\% on VoxCeleb1-O and also brings overall improvements on the remaining test splits. MHC-ResNet34 also outperforms the ResNet-34 baseline on VoxCeleb1, indicating that mHC can benefit CNN-based speaker encoders as well. For ECAPA-TDNN, the improvements are more pronounced: MHC-ECAPA-S achieves 0.98\%/1.07\%/2.06\% EER on O/E/H versus 1.02\%/1.41\%/2.26\% for the baseline, and MHC-ECAPA-L further reduces EER from 0.87\%/1.12\%/2.12\% to 0.77\%/0.94\%/1.88\% on O/E/H, with reductions in the corresponding MinDCF. Notably, MHC-ECAPA-L achieves 0.77\% EER on VoxCeleb1-O, 0.94\% on VoxCeleb1-E, and 1.88\% on VoxCeleb1-H. These results are highly competitive, demonstrating that mHC can further strengthen a strong ECAPA-TDNN-L baseline.

To further validate the effectiveness of mHC under a more challenging evaluation condition, we additionally report results on VoxSRC21-val in Table~\ref{tab:voxsrc21_small}. Compared with their corresponding baselines, mHC variants achieve consistent reductions in both EER and MinDCF, with the largest relative gain observed for ResNet-34 (EER 3.83\% $\rightarrow$ 3.35\%). These results demonstrate that the benefit of mHC generalizes beyond VoxCeleb1 and remains effective on the VoxSRC21-val benchmark.

\begin{table*}[t]
\centering
\caption{EER and MinDCF performance of all systems on the standard VoxCeleb1 test sets. $\triangle$(\%) is the relative reduction in EER.}
\label{tab:voxceleb_results}
\renewcommand{\arraystretch}{0.9}
\resizebox{\textwidth}{!}{
\footnotesize 
\setlength{\tabcolsep}{4pt} 
\begin{tabular}{lc|ccc|ccc|ccc}
\toprule
\textbf{Architecture} & \textbf{\# Params} & \multicolumn{3}{c}{\textbf{VoxCeleb1-O}} & \multicolumn{3}{c}{\textbf{VoxCeleb1-E}} & \multicolumn{3}{c}{\textbf{VoxCeleb1-H}}  \\
&(M) & \textbf{EER(\%)} & \textbf{$\triangle$(\%)} & \textbf{MinDCF} & \textbf{EER(\%)} & \textbf{$\triangle$(\%)} & \textbf{MinDCF} & \textbf{EER(\%)} & \textbf{$\triangle$(\%)} & \textbf{MinDCF} \\
\midrule
Res2Net \cite{gao2019res2net} & 4.03 & 1.56 & - & \textbf{0.150} & 1.41 & - & 0.149 & 2.48 & - & 0.230 \\
\rowcolor{myyellow}
\textbf{mHC-Res2Net} & 4.03 &\textbf{1.41}  & \cellcolor{mygreen}+9.6 &0.154  &\textbf{1.40}  & \cellcolor{mygreen}+0.7 &\textbf{0.145}  &\textbf{2.43}   & \cellcolor{mygreen}+2.0 &\textbf{0.228}  \\
\midrule
ResNet34 \cite{he2016deep} & 6.34 & 1.05 & - & 0.107 & 1.11 & - & 0.116 & 1.99 & - & 0.192 \\
\rowcolor{myyellow}
\textbf{mHC-ResNet34} & 6.34 & \textbf{1.03} & \cellcolor{mygreen}+1.9 & \textbf{0.095} & \textbf{1.09} & \cellcolor{mygreen}+1.8 & \textbf{0.110} & \textbf{1.93} & \cellcolor{mygreen}+3.0 & \textbf{0.188} \\
\midrule
ECAPA-TDNN-S \cite{desplanques2020ecapa} & 6.19 & 1.02 & - & 0.106 & 1.41 & - & 0.162 & 2.26 & - & 0.257 \\
\rowcolor{myyellow}
\textbf{mHC-ECAPA-S} & 6.19 & \textbf{0.98} &\cellcolor{mygreen}+3.9  &0.107  &\textbf{1.07}  &\cellcolor{mygreen}+24.1  &0.119  &\textbf{2.06} &\cellcolor{mygreen}+8.8 &\textbf{0.204} \\
\midrule
ECAPA-TDNN-L \cite{desplanques2020ecapa} & 20.76 & 0.87 & - & 0.107 & 1.12 & - & 0.132 & 2.12 & - & 0.210 \\
\rowcolor{myyellow} 
\bf{mHC-ECAPA-L} & 20.76 & \bf{0.77} & \cellcolor{mygreen}+11.5 & \bf{0.084} & \bf{0.94} & \cellcolor{mygreen}+16.1 & \bf{0.110} & \bf{1.88} & \cellcolor{mygreen}+11.3 & \bf{0.190} \\
\bottomrule
\end{tabular}
}
\end{table*}

\begin{table}[t] 
\centering
\caption{Performance on VoxSRC21-val. $\triangle$(\%) is the relative reduction in EER.}
\label{tab:voxsrc21_small}
\renewcommand{\arraystretch}{0.9} 
\resizebox{\columnwidth}{!}{ 
\footnotesize 
\setlength{\tabcolsep}{4pt} 
\begin{tabular}{lc|ccc}
\toprule
\textbf{Architecture} & \textbf{Params} & \multicolumn{3}{c}{\textbf{VoxSRC21-val}} \\
& (M) & \textbf{EER(\%)} & \textbf{$\triangle$(\%)} & \textbf{MinDCF} \\
\midrule
Res2Net \cite{gao2019res2net} & 4.03 & 4.27 & - & 0.334 \\
\rowcolor{myyellow}
\textbf{mHC-Res2Net} & 4.03 & \textbf{4.13} & \cellcolor{mygreen}+3.3 & \textbf{0.332} \\
\midrule
ResNet34 \cite{he2016deep} & 6.34 & 3.83 & - & 0.316 \\
\rowcolor{myyellow}
\textbf{mHC-ResNet34} & 6.34 & \textbf{3.35} & \cellcolor{mygreen}+12.5 & \textbf{0.269} \\
\midrule
ECAPA-TDNN-S \cite{desplanques2020ecapa} & 6.19 & 4.26 & - & 0.384 \\
\rowcolor{myyellow}
\textbf{mHC-ECAPA-S} & 6.19 & \textbf{4.11} & \cellcolor{mygreen}+3.5 & \textbf{0.324} \\
\midrule
ECAPA-TDNN-L \cite{desplanques2020ecapa} & 20.76 & 3.89 & - & 0.323 \\
\rowcolor{myyellow} 
\textbf{mHC-ECAPA-L } & 20.76 & \textbf{3.68} & \cellcolor{mygreen}+5.4 & \textbf{0.316} \\
\bottomrule
\end{tabular}
}
\end{table}


\subsection{Ablation Study}
\begin{figure}[!t]
  \centering
  \includegraphics[width=0.6\columnwidth]{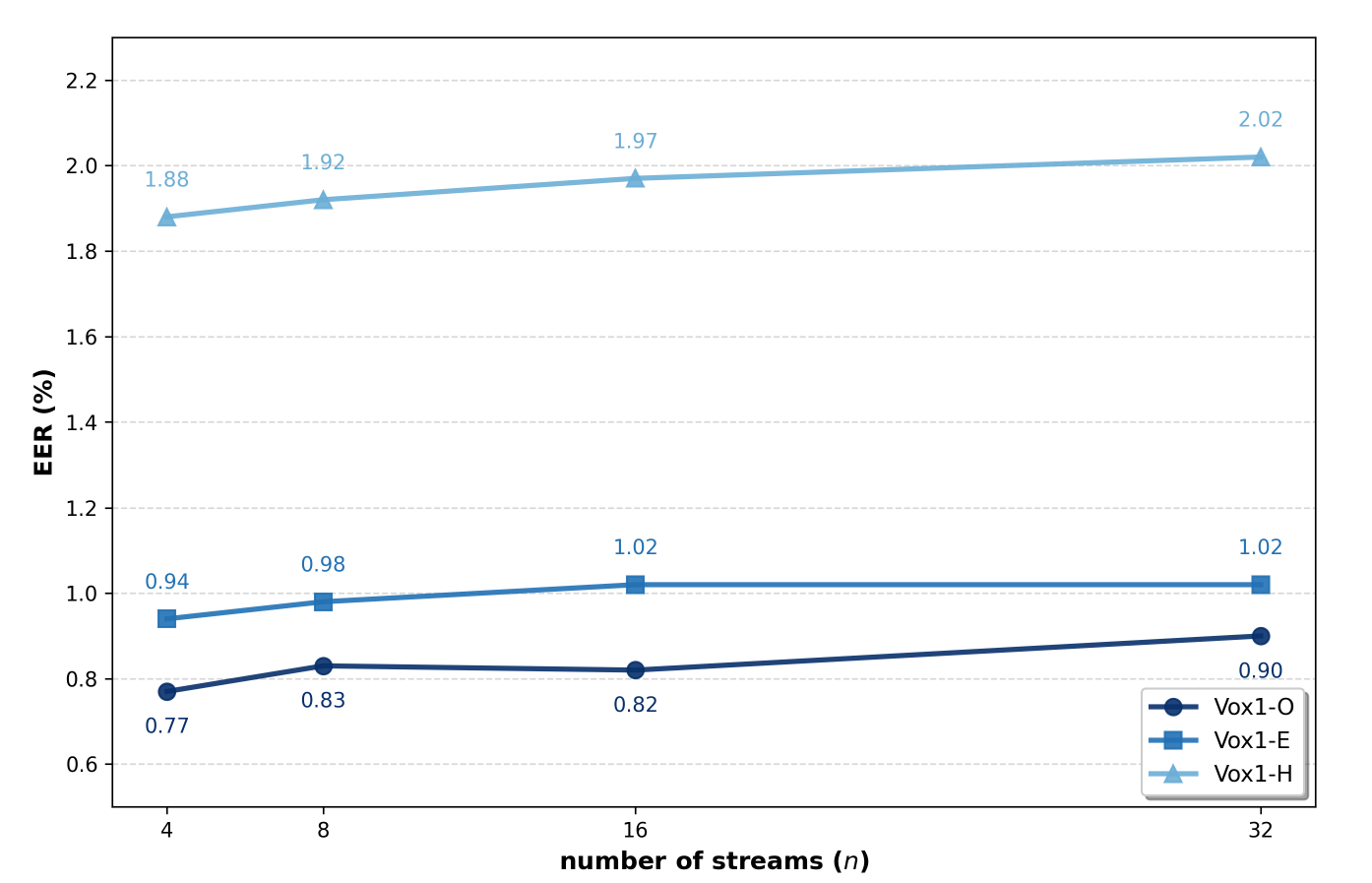}
  \caption{Effect of the number of parallel streams $N$ in mHC.}
  \label{fig:numstreams_analysis}
\end{figure}

We investigate the influence of the number of parallel streams $N$ on performance using the ECAPA-TDNN-L architecture. Experiments are conducted across $N \in \{4, 8, 16, 32\}$ and evaluated on the VoxCeleb1-O, E, and H test sets. As illustrated in Fig.~\ref{fig:numstreams_analysis}, the EER is consistently the lowest at $N=4$ for all three evaluation sets. We observe a marginal increase in EER as the value of $N$ increases, suggesting that a smaller stream count provides a more effective trade-off between cross-stream information exchange and model convergence stability.

\begin{table}[!t]
  \centering
  
  \caption{Comparison between mHC and HC on VoxCeleb1-O (ECAPA-TDNN-L).}
  \label{tab:mhc_vs_hc}
  \renewcommand{\arraystretch}{0.9}
  \setlength{\tabcolsep}{6pt}
  \begin{tabular}{l c}
    \toprule
    \textbf{Method} & \textbf{EER(\%)} \\
    \midrule
    HC & 0.84 \\
    \bf{mHC} & \textbf{0.77} \\
    \bottomrule
  \end{tabular}
\end{table}

To verify the effectiveness of the manifold constraint, we conduct a comparison between HC and mHC on ECAPA-TDNN-L, as reported in Table~\ref{tab:mhc_vs_hc}. The manifold-constrained design reduces the EER from 0.84\% to 0.77\%, confirming that enforcing the doubly stochastic manifold is effective.


\subsection{Efficiency and Complexity Analysis}
To demonstrate that mHC achieves superior performance without imposing significant overhead, we conduct an efficiency and complexity analysis comparing the baseline models with their mHC-enhanced versions across two scales: ECAPA-TDNN-S and ECAPA-TDNN-L.

\begin{figure}[t]
  \centering
  \includegraphics[width=0.75\columnwidth]{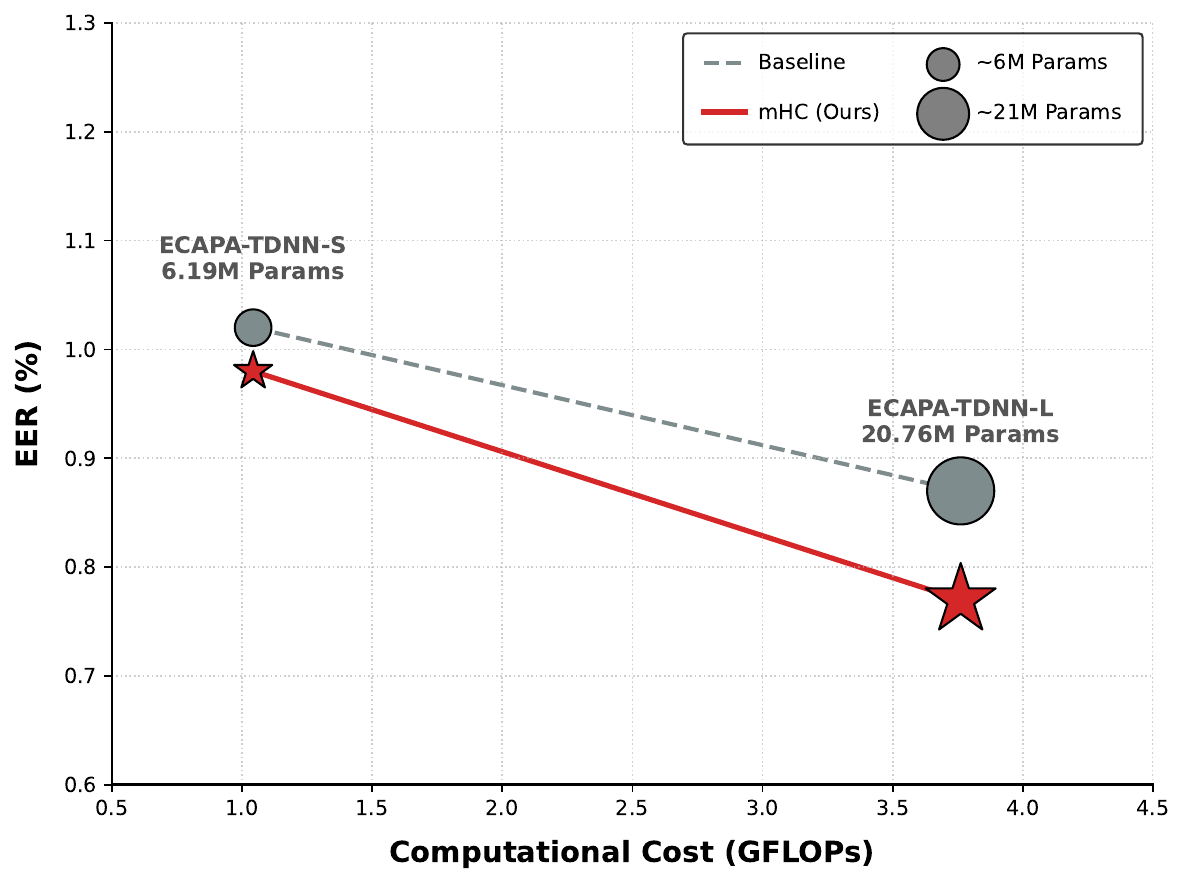}
  \caption{Computational cost (GFLOPs) comparison between baseline backbones and their mHC variants.}
  \label{fig:mhc_flops}
\end{figure}

As illustrated in Fig.~\ref{fig:mhc_flops}, the introduction of the manifold-constrained interaction mechanism results in negligible changes to both the parameter count and computational cost. For both the Small (S) and Large (L) variants, the mHC versions maintain nearly identical GFLOPs—approximately 1.0 GFLOPs and 3.8 GFLOPs respectively—while achieving a substantial reduction in equal error rate (EER) across all evaluation sets. This indicates that the mixing matrix $\mathbf{W}$ and the associated Sinkhorn-Knopp iterations add minimal floating-point operations relative to the backbone’s total computation. Consequently, mHC serves as a highly efficient architectural component that enhances representational power without compromising the real-time processing capabilities required for practical speaker recognition systems.

\section{Conclusions}
In this paper, we investigated the application of Manifold-Constrained Hyper-Connections (mHC) to enhance speaker recognition systems. By integrating mHC into ResNet, Res2Net, and ECAPA-TDNN, we demonstrated that replacing conventional residual shortcuts with manifold-constrained multi-stream mixing improves speaker embeddings while preserving stable signal propagation through a doubly stochastic matrix. Experiments on VoxCeleb1 and VoxSRC21-val show consistent EER and MinDCF reductions across multiple backbone architectures with negligible parameter and FLOP overhead. The comparison between HC and mHC further confirms the importance of the manifold constraint.

\newpage
\section{Acknowledgement} This work was supported in part by the Research Grants Council of the Hong Kong SAR (Grant No. 15228223), and The Hong Kong Polytechnic University, Project ID P0049192.

\section{Generative AI Use Disclosure} Generative AI tools were used only for language polishing and formatting assistance. All scientific content, experiments, analyses, and conclusions were produced and verified by the authors.

\begingroup
\renewcommand{\eightpt}{\fontsize{8.0}{8.2}\selectfont}
\bibliographystyle{IEEEtran}
\bibliography{mybib}
\endgroup

\end{document}